\documentclass[12pt,a4paper]{article}
\usepackage{amsmath, amssymb}
\usepackage{amsfonts}
\usepackage[mathscr]{eucal}
\usepackage{ascmac}
\usepackage[dvips]{graphicx}
\usepackage{subfigure}
\usepackage[charter]{mathdesign}
\usepackage{bm}
\usepackage{cite}
\usepackage{here}
\usepackage{color}
\usepackage{comment}

\usepackage[height=22.5cm,width=16.5cm]{geometry}
\begin{document}
\renewcommand{\thefootnote}{$\clubsuit$\arabic{footnote}}
\def\a{\alpha}
\def\b{\beta}
\def\c{\varepsilon}
\def\d{\delta}
\def\e{\epsilon}
\def\f{\phi}
\def\g{\gamma}
\def\h{\theta}
\def\k{\kappa}
\def\l{\lambda}
\def\m{\mu}
\def\n{\nu}
\def\p{\psi}
\def\q{\partial}
\def\r{\rho}
\def\s{\sigma}
\def\t{\tau}
\def\u{\upsilon}
\def\v{\varphi}
\def\w{\omega}
\def\x{\xi}
\def\y{\eta}
\def\z{\zeta}
\def\D{\Delta}
\def\G{\Gamma}
\def\H{\Theta}
\def\L{\Lambda}
\def\F{\Phi}
\def\P{\Psi}
\def\S{\Sigma}

\def\der{\partial}
\def\o{\over}
\def\beq{\begin{eqnarray}}
\def\eeq{\end{eqnarray}}
\newcommand{\gsim}{ \mathop{}_{\textstyle \sim}^{\textstyle >} }
\newcommand{\lsim}{ \mathop{}_{\textstyle \sim}^{\textstyle <} }
\newcommand{\vev}[1]{ \left\langle {#1} \right\rangle }
\newcommand{\bra}[1]{ \langle {#1} | }
\newcommand{\ket}[1]{ | {#1} \rangle }
\newcommand{\EV}{ {\rm eV} }
\newcommand{\KEV}{ {\rm keV} }
\newcommand{\MEV}{ {\rm MeV} }
\newcommand{\GEV}{ {\rm GeV} }
\newcommand{\TEV}{ {\rm TeV} }
\def\diag{\mathop{\rm diag}\nolimits}
\def\Spin{\mathop{\rm Spin}}
\def\SO{\mathop{\rm SO}}
\def\O{\mathop{\rm O}}
\def\SU{\mathop{\rm SU}}
\def\U{\mathop{\rm U}}
\def\Sp{\mathop{\rm Sp}}
\def\SL{\mathop{\rm SL}}
\def\tr{\mathop{\rm tr}}
\def\Dterm{_{\theta^2\tilde{\theta^2}}}
\def\Fterm{_{\theta^2}}
\def\F*term{_{\tilde{\theta}^2}}
\def\mi{m_{\phi}}
\def\mpl{M_{\rm pl}}

\def\IJMP{Int.~J.~Mod.~Phys. }
\def\MPL{Mod.~Phys.~Lett. }
\def\NP{Nucl.~Phys. }
\def\PL{Phys.~Lett. }
\def\PR{Phys.~Rev. }
\def\PRL{Phys.~Rev.~Lett. }
\def\PTP{Prog.~Theor.~Phys. }
\def\ZP{Z.~Phys. }


\makeatletter
\@addtoreset{equation}{section}
\def\theequation{\thesection.\arabic{equation}}
\makeatother
\newcommand{\Slash}[1]{{\ooalign{\hfil/\hfil\crcr$#1$}}} 

\newcommand{\TODO}[1]{{$[[ \clubsuit\clubsuit$ \bf #1 $\clubsuit\clubsuit ]]$}}
\newcommand{\kmrem}[1]{{\color{red} \bf $[[ $ KM: #1$ ]]$}}
\newcommand{\knrem}[1]{{\color{blue} \bf $[[ $ KN: #1$ ]]$}}
\newcommand{\mtrem}[1]{{\color{green} \bf $[[ $ MT: #1$ ]]$}}
\newcommand{\rem}[1]{{\color{red} \bf $[[ $#1$ ]]$}}

\begin{titlepage}

\begin{flushright}
UT-14-15
\end{flushright}

\vskip 3cm

\begin{center}
{
\Huge 
\bfseries
Dark Matter Chaotic Inflation \\[1em]
in Light of BICEP2
}

\vskip .65in
{\large
Kyohei Mukaida$^{\spadesuit}$ and 
Kazunori Nakayama$^{\spadesuit,\diamondsuit}$
}

\vskip .35in
\begin{tabular}{ll}
$^{\spadesuit}$ &\!\! {\em Department of Physics, Faculty of Science, }\\
& {\em University of Tokyo,  Bunkyo-ku, Tokyo 133-0033, Japan}\\[.5em]
$^{\diamondsuit}$ &\!\! {\em Kavli Institute for the Physics and Mathematics of the Universe, }\\
&{\em Todai Institute for Advanced Study,}\\
&{\em University of Tokyo,  Kashiwa, Chiba 277-8583, Japan}
\end{tabular}

\vskip .65in

\begin{abstract}
We propose an economical model in which a singlet Z$_2$-odd scalar field accounts for
the primordial inflation and the present dark matter abundance simultaneously
in the light of recent BICEP2 result.
Interestingly, the reheating temperature and the thermal dark matter abundance are closely connected
by the same interaction between the singlet scalar and the standard model Higgs.
In addition, the reheating temperature turns out to be quite high, $T_\text{R} \gtrsim 10^{12}\,\GEV$,
and hence the thermal leptogenesis is compatible with this model.
Therefore, it can be one of the simplest cosmological scenarios.
\end{abstract}

\end{center}

\end{titlepage}

\newpage


\section{Introduction and Summary}
\label{sec:}

Recently, the BICEP2 experiment discovered the B-mode polarization in the cosmic microwave background (CMB)
anisotropy, which is interpreted as the primordial gravitational waves of the inflationary origin~\cite{Ade:2014xna}.
This confirms the idea of inflation~\cite{Guth:1980zm,Linde:1981mu}, especially the high scale inflation
such as the chaotic inflation~\cite{Linde:1983gd}.

On the other hand, one of the greatest mysteries of the Universe is the presence of dark matter (DM)~\cite{Ade:2013rta}.
Since there is no candidate for the DM in the standard model (SM) of particle physics,
it clearly requires physics beyond the SM.
Maybe the simplest extension of the SM is to add a singlet scalar field $\phi$ which has a Z$_2$-symmetry~\cite{Silveira:1985rk,McDonald:1993ex} and couples to the SM Higgs boson $H$ in the scalar potential as
\begin{equation}
	V = \frac{1}{2}m_\phi^2\phi^2 + \frac{1}{2}g^2\phi^2|H|^2,
\end{equation}
where $g$ is a coupling constant.
Due to the Z$_2$-symmetry under which $\phi$ transforms as $\phi \to -\phi$, it is stable.
It can have a correct annihilation cross section through the Higgs portal for $m_\phi \sim \mathcal O(100)$\,GeV
and $g \sim \mathcal O(1)$ and account for the observed amount of DM.

We show that the scalar singlet DM, $\phi$, can cause inflation which is consistent with the BICEP2 result.
Naively, at the large field value, $\phi$ obtains a $\phi^4$ potential radiatively,
and hence this chaotic inflation with $\phi^4$ potential is already ruled out.
Moreover, it is difficult  to account for the observed density perturbation of the Universe.
Our idea is to modify the kinetic term of $\phi$ so that the potential becomes quadratic in terms of the canonically normalized field.
This is the so-called running kinetic inflation~\cite{Nakayama:2010kt,Nakayama:2010sk}.
It has been shown that the SM Higgs boson can be the inflaton to be consistent with the BICEP2 result~\cite{Nakayama:2014koa}.

In this paper, we identify the singlet scalar $\phi$ as the inflaton, and show that it can simultaneously explain the present DM abundance.
The key feature is that the inflaton coherent oscillation can be soon dissipated away
even if it is perturbatively stable at its vacuum,
and eventually the inflaton itself participates in the thermal plasma
as extensively studied in Refs.~\cite{Mukaida:2012qn,Mukaida:2013xxa,Mukaida:2014yia}.
Then, as the Universe expands, the annihilation of the inflaton particles again decouples from the thermal plasma
at late time,
which leads to the standard freeze-out DM production.
This scenario typically results in a high reheating temperature $T_\text{R} \gtrsim 10^{12}\, \GEV$ 
that is compatible with the thermal leptogenesis~\cite{Fukugita:1986hr}.
Interestingly, the coupling $g$, which determines the present relic DM abundance,
also determines the reheating temperature of the Universe.
In this sense, the model we propose is quite economical:\footnote{
	A similar model was proposed in Refs.~\cite{Lerner:2009xg,Okada:2010jd} 
	in the context of inflation with non-minimal
	coupling to gravity~\cite{Bezrukov:2007ep}.
	In Ref.~\cite{Nakayama:2010kt}, the possibility of inflatino DM (the supersymmetric partner of the inflaton) 
	in the context of running kinetic inflation was pointed out.
}
we can explain the inflation, reheating and DM consistent with observations by just adding a real scalar $\phi$.

\section{Singlet Dark Matter as Inflaton}
\label{sec:}
In this section, let us consider a singlet scalar field $\phi$ which is a Z$_2$-odd state,
while the other SM 
fields are Z$_2$-even states.
By imposing the Z$_2$ invariance to the action,
this singlet particle $\phi$ becomes stable and can be a candidate of DM. 
The only renormalizable coupling of $\phi$ with SM fields is a quartic interaction with the SM Higgs doublet:
$g^2 \phi^2 |H|^2$.
The coupling $g$ and the mass $m_\phi$ are constrained by recent observations~\cite{Cline:2013gha}.

Since this term inevitably introduces a four point interaction of $\phi$,
we need some modifications to the potential of $\phi$ in the light of recent constraints on
the models of the inflation.
Thus, let us embed this scenario into the running kinetic inflation~\cite{Nakayama:2010kt} in the following,
which results in the quadratic chaotic inflation favored by the BICEP2 experiment.

\subsection{Setup}
\label{sec:}
Suppose a singlet scalar field $\phi$ with a potential invariant under the Z$_2$-symmetry $\phi \to - \phi$.
As stated above, since the quartic chaotic inflation is already excluded,
we need some modifications to the potential of $\phi$.
To account for the flatter potential which is required for an observationally favored inflation,
we impose a shift symmetry $\phi^2 \to \phi^2 + C$ where $C$ is a real parameter.
This symmetry is assumed to be broken at the low energy scale explicitly.
Then, the relevant Lagrangian can be written as\footnote{
	In this paper, we do not consider higher derivative terms of $\phi^2$ which
	also respect the shift symmetry.
}
\begin{align}
\label{eq:setup}
	{\cal L} = \frac{1}{2} \left( \frac{\der_\mu \phi^2}{2\mpl} \right)^2
	+ \epsilon \frac{1}{2} \left( \der_\mu \phi \right)^2 - \frac{1}{2} m_\phi^2 \phi^2 
	- \frac{1}{4} \lambda \phi^4 - \frac{1}{2} g^2 \phi^2 |H|^2 + \cdots,
\end{align}
at the lowest order.
Here $\mpl$ stands for the reduced Planck mass.
The first term respects the shift symmetry, while the other terms break it very weakly:
$\epsilon, \lambda, g^2, m_\phi/\mpl \ll 1$.
Notice that the second term is responsible for the kinetic term at the low energy,
while the first term gives the kinetic term at the inflationary (large field value) regime.
To see this behavior more explicitly, let us canonically normalize the inflaton field $\phi$.
Since the Lagrangian \eqref{eq:setup} reads
\begin{align}
	{\cal L} = \frac{1}{2} \left( \frac{\phi^2}{\mpl^2} + \epsilon \right) \left( \der_\mu \phi \right)^2
	- \frac{1}{2} m_\phi^2 \phi^2 
	- \frac{1}{4} \lambda \phi^4 - \frac{1}{2} g^2 \phi^2 |H|^2 + \cdots,
\end{align}
one obtains the canonically normalized field $\sigma$ as
\begin{align}
	\frac{\sigma}{\e \mpl} = \frac{1}{2} \frac{\phi}{\sqrt{\e} \mpl} \sqrt{\frac{\phi^2}{\e \mpl^2} + 1}
	+\frac{1}{2} \ln \left[  \frac{\phi}{\sqrt{\e} \mpl} + \sqrt{\frac{\phi^2}{\e \mpl^2} + 1} \right].
\end{align}
This analytic form can be approximated with
\begin{align}
\label{eq:can_field}
	\frac{\sigma}{\e \mpl} \simeq
	\begin{cases}
		\cfrac{\phi}{\sqrt{\e} \mpl} +\cfrac{1}{6} \left( \cfrac{\phi}{\sqrt{\e} \mpl} \right)^{3} + \cdots
		&\text{for}~~|\phi| \ll \sqrt{\e} \mpl , \\[20pt]
		\pm\cfrac{1}{2} \cfrac{\phi^2}{\e \mpl^2} 
		\pm \cfrac{1}{4} \left[ 1 + 2 \ln \left( \cfrac{2 |\phi|}{\sqrt{\e} \mpl} \right) \right] + \cdots
		&\text{for}~~|\phi| \gg \sqrt{\e} \mpl ,
	\end{cases}
\end{align}
where the sign in the second line corresponds to the sign of $\sigma$.

Before discussing the inflationary dynamics,
let us relate this action with one at the low energy scale.
From Eq.~\eqref{eq:can_field}, the canonically normalized field $\sigma$ for a small field value $\phi \ll \sqrt{\epsilon} \mpl$
becomes $\sigma \simeq \sqrt{\epsilon} \phi  \equiv \tilde \phi$.
Thus, the low energy effective action can be expressed as
\begin{align}
	{\cal L} = \frac{1}{2} \left( \der_\mu \tilde \phi \right)^2 - \frac{1}{2} \tilde m_\phi^2 \tilde \phi^2
	-\frac{1}{4} \tilde \lambda \tilde \phi^4 - \frac{1}{2} \tilde g^2 \tilde \phi^2 |H|^2 + \cdots ,
\end{align}
where the parameters at the transition scale, $\epsilon \mpl$, are defined as
$\tilde m^2_\phi \equiv m_\phi^2/\epsilon$, $\tilde \lambda \equiv \lambda/\epsilon^2$ and 
$\tilde g^2 \equiv g^2/ \epsilon$.
Assuming that the self interaction is smaller than the quartic interaction at $\tilde m_\phi$ scale,
$\tilde \lambda \ll \tilde g^4$,
one finds that the radiative correction dominates the self interaction of $\phi$ at the high energy scale.\footnote{
	One may regard $g^2$ as an order parameter of the shift symmetry breaking.
	Then we naturally expect $\lambda \sim g^4$. The presence of $\lambda$ of this order does not much affect the following results.
}
Hence, we expect the following relation:\footnote{
	Also the quartic interaction increases the Higgs four point interaction radiatively,
	and hence it can stabilize the Higgs potential.
}
\begin{align}
\label{eq:lambda_ir}
	\lambda/ \epsilon^2 = \tilde \lambda \simeq (\tilde g^4 / 8 \pi^2) \ln [\epsilon \mpl/\tilde m_\phi].
\end{align}

The important low energy parameters $\tilde g$ and $\tilde m_\phi$ are being constrained by
recent observations~\cite{Cline:2013gha}.
There are basically two viable parameter regions: the light singlet region and heavy singlet region,
in which the observed DM abundance is correctly explained with satisfying other experimental constraints.
First, in the light singlet region: $[\tilde m^2_\phi + \tilde \lambda^2 v^2 /2]^{1/2} < m_h/2$\footnote{
	The physical mass of DM, $[\tilde m^2_\phi + \tilde \lambda^2 v^2 /2]^{1/2}$, can be approximated with $\tilde m_\phi$
	because the coupling at $\tilde m_\phi$ scale $\tilde \lambda (\ll \tilde g^4)$
	turns out to be small in the viable parameter regions.
	See Eq.~\eqref{eq:bmp}.
}, 
where $v \simeq 246\, \GEV$ and $m_h \simeq 125\, \GEV$ are the VEV and mass of the SM Higgs respectively,
the couplings should be smaller than $\tilde g^2  \lesssim 0.02$--$0.03$
owing to the constraint on the invisible decay width of the Higgs boson.
In this regime, since the coupling is somewhat small,
the DM mass should be near the resonance pole of the Higgs boson to account for the correct DM abundance.
Hence, the allowed region lies near 
$\tilde m_\phi \sim m_h/2$.
For the heavy singlet case, the DM direct detection experiments give stringent constraint on the parameter spaces.
Notice that a too large mass requires a large coupling to account for the present DM abundance,
and it eventually threatens the perturbativity of this model.

As reference values, we take following parameters:
\begin{equation}
\label{eq:bmp}
\begin{cases}
	\tilde m_\phi = 55\,{\rm GeV},~~\tilde g ^2 = 4\times 10^{-3} &{\rm for~light~singlet~case},\\
	\tilde m_\phi = 1\,{\rm TeV},~~\tilde g^2 = 0.3 &{\rm for~heavy~singlet~case}.
\end{cases}
\end{equation}
These parameter regions may be probed by future studies of Higgs invisible invisible decay width
at ILC~\cite{Peskin:2012we}
or the DM direct detection such as XENON1T~\cite{Aprile:2012zx}.

\subsection{Running kinetic inflation with singlet scalar DM}
\label{sec:}
Then, let us discuss the inflationary dynamics of this model.
For a large field value $\phi \gg \sqrt{\epsilon} \mpl$, the relevant Lagrangian \eqref{eq:setup} in terms of a
canonically normalized field $\sigma \simeq \pm\phi^2 / (2 \mpl) \equiv \hat \phi$
can be written as
\begin{align}
\label{eq:inf_regime}
	{\cal L} = \frac{1}{2} \left( \der_\mu \hat \phi \right)^2 - \lambda \mpl^2 \hat \phi^2 + \cdots,
\end{align}
where $\cdots$ collectively denotes small terms which are not relevant in the inflationary regime
as justified a posteriori.
Thus, this model can account for the quadratic chaotic inflation at the large field value $|\hat\phi| \gg M_{\rm pl}$.
The effective inflaton mass in this regime is given by $M^2 = 2 \lambda \mpl^2$.

To satisfy the Planck normalization of the scalar fluctuations,
the mass of the inflaton should be $M \simeq 1.5\times10^{13}\, \GEV$.
This fixes the coupling $\lambda$ as
\begin{align}
\label{eq:lambda}
	\lambda = \frac{M^2}{2 \mpl^2} \simeq 2\times 10^{-11}.
\end{align}
Then, Eq.~\eqref{eq:lambda_ir} implies that the small parameter $\epsilon$ is roughly given by
\begin{align}
\label{eq:epsilon}
	\epsilon \sim \frac{8\times 10^{-6}}{\tilde g^2}
	\sim \begin{cases}
		2\times 10^{-5} &{\rm for~light~singlet~case}, \\
		2\times 10^{-3} &{\rm for~heavy~singlet~case}.
	\end{cases}
\end{align}
Hence the transition field value becomes $\epsilon \mpl \sim 10^{13-15}\,\GEV$.
By using Eq.~\eqref{eq:epsilon}, we can estimate typical values of original ``small'' parameters $m_\phi, g^2$.
Interestingly, the Planck normalization fixes the combination, $\e \tilde g^2$, and hence one finds
\begin{align}
	g^2 & \sim 8 \times 10^{-6}.
\end{align}
Also the typical value of the inflaton mass should be
\begin{align}
	m_\phi=\sqrt{\epsilon} \tilde m_\phi \sim \mathcal O(1)\, \GEV,
\end{align}
for both the light singlet and heavy singlet cases.
Since these parameters are small,
the inflationary dynamics can be well described by Eq.~\eqref{eq:inf_regime}.
Also, notice that the field direction along the Higgs field acquires larger mass  than the Hubble scale
during the inflationary regime: $g\sqrt{\hat\phi_\text{inf}M_{\rm pl}} \sim 10^{15} \,\GEV > H_\text{inf} \sim 10^{14}\, \GEV$
where $H_{\rm inf}$ denotes the Hubble scale during inflation.
Therefore, the Higgs field is expected to settle into its vacuum immediately.

\subsection{Reheating after Inflation}
\label{sec:}
After the era of the inflation,
the inflaton $\phi$ starts to oscillate coherently around its potential minimum
with a large initial amplitude $\sim \mpl$.
Since the Higgs field couples to the inflaton via the quartic coupling,
its effective mass term depends on the field value of $\phi$, and 
the dispersion relation of the Higgs depends on time in the inflaton oscillation regime.
Hence, the Higgs particles can be produced copiously via the so-called non-perturbative particle production~\cite{Kofman:1994rk}.
Moreover, the Higgs has a large top Yukawa coupling, 
and produced Higgs particles can decay into other SM particles efficiently,
that is, so-called the instant preheating~\cite{Felder:1998vq} takes place.
In this case, the subsequent reheating stage can be roughly divided into two regimes;
(i) the instant preheating creates a (hot) background plasma
and it terminates due to efficient rescatterings,
(ii) then the remaining inflaton condensation dissipates its energy
via the interaction with the background plasma~\cite{Yokoyama:2004pf,BasteroGil:2010pb,Drewes:2010pf}.
Complete analyses of this whole reheating process are found in a series of works~\cite{Mukaida:2012qn,Mukaida:2013xxa,Mukaida:2014yia}.

To estimate the reheating temperature in such a complicated scenario,
we roughly approximate the potential in terms of the canonically normalized field $\sigma$ with
\begin{align}
	V \simeq 
	\begin{cases}
		\cfrac{1}{2}M^2 \sigma^2 +  \e \tilde g^2 \mpl |\sigma| |H|^2
		&\text{for}~~|\sigma| > \e \mpl, \\[15pt]
		\cfrac{1}{2} \tilde m_\phi^2 \sigma^2 + \cfrac{1}{4} \tilde \lambda \sigma^4
		+ \cfrac{1}{2} \tilde g^2 \sigma^2 |H|^2
		&\text{for}~~|\sigma| < \e \mpl,
	\end{cases}
\end{align}
where $M^2/2 \equiv \e^2 \tilde \lambda \mpl^2$ and $\tilde \lambda \sim \tilde g^4/8\pi^2$.
The inflaton obeys the following equation of motion:
\begin{align}
	\ddot \sigma +3 H \dot\sigma + V' (\sigma) + \Gamma_\sigma [\sigma; T] \dot \sigma = 0,
\end{align}
where the dissipation rate is denoted as $\Gamma_\sigma [\sigma; T]$ that
depends on the field value $\sigma$ and also the property of background plasma in general.
Here we simply assumed that the background plasma can be approximated with the thermal one.
The evaporation time of the inflaton condensation is characterized by the oscillation averaged dissipation rate~\cite{Mukaida:2012qn}:
\begin{align}
	\Gamma_\sigma^\text{eff} [\bar \sigma; T] 
	\equiv \frac{\left< \Gamma_\sigma [\sigma; T] \dot \sigma^2 \right>}{\left< \dot \sigma^2 \right>},
\end{align}
where $\left< \bullet \right>$ denotes the oscillation time average.
By using the effective dissipation rate,
the evolution of the Universe can be obtained from following equations:
\begin{align}
\label{eq:eom1}
	\dot \rho_\sigma & = - \left[ x_\sigma H + y_\sigma \Gamma_\sigma^\text{eff} \right] \rho_\sigma, \\[10pt]
\label{eq:eom2}
	\dot \rho_\text{rad} & = - 4 H \rho_\text{rad} + y_\sigma \Gamma_\sigma^\text{eff} \rho_\sigma, \\[10pt]
\label{eq:eom3}
	3 \mpl^2 H ^2 &= \rho_\sigma + \rho_\text{rad},
\end{align}
where $\rho_{\sigma/ \text{rad}}$ denotes the energy density of the inflaton/radiation respectively
and $(x_\sigma, y_\sigma)$ depends on which term dominates the potential: 
$(x_\sigma, y_\sigma) = (3,1)$ for the quadratic term and $(x_\sigma, y_\sigma)= (4,4/3)$ for the quartic term.
Since the only unknown function is $\Gamma_\sigma^\text{eff}$,
let us evaluate this effective dissipation rate in the following.

At first, the potential is dominated by the large field value regime, 
since its amplitude $\bar \sigma$ is large, $\bar \sigma > \e \mpl$.
Thus, the inflaton approximately oscillates with $\sigma = \bar\sigma \cos [M t]$,
and hence the velocity of the inflaton field around its origin is given by
$\left. \dot \sigma \right|_{\sigma \sim 0} \simeq M \bar \sigma$.
On the other hand,
after its amplitude decreases due to the cosmic expansion, 
its potential becomes dominated by the four point interaction.
Then the velocity reads $\left. \dot \sigma \right|_{\sigma \sim 0} \simeq \tilde\lambda^{1/2} \bar\sigma^2$.
Owing to this large velocity, the adiabaticity of the Higgs can be broken down near the origin of the inflaton potential
and then they may be produced non-perturbatively.
Let us define a parameter $k_\ast$ that characterizes the non-perturbative particle production as
\begin{align}
\label{eq:kstar}
	k_\ast^2 \equiv \tilde g \left. \dot\sigma \right|_{\sigma \sim 0}.
\end{align}
The Higgs particles are efficiently produced if the following condition is met~\cite{Kofman:1994rk}:
\begin{align}
\label{eq:np_cond}
	k_\ast^2 \gg \max \left[ m_{\text{eff},\sigma}^2,  m_{\text{scr},H}^2 \right],
\end{align}
where $m_{\text{eff},\sigma}$ is the effective mass of the inflaton that depends on which term
dominates the potential, and
$m_{\text{scr},H}$ denotes the screening mass of the Higgs near the potential origin of the inflaton,
which comes from interactions with the background plasma if it exists.
If the background plasma can be regarded as the thermal one,
the screening mass for the Higgs is approximated with $y_t T$.
Before the first passage of $\sigma \sim 0$, since there is no background plasma in the Universe,
this screening mass vanishes.
The first condition implies that the non-perturbative particle production does not take place
if the mass of the inflaton is larger than the effective mass of the Higgs, $m_{\text{eff}, \sigma} > \tilde g \bar\sigma$;
and the second one implies that if the interaction between the Higgs and the background plasma is efficient,
the non-perturbative particle production does not occur.
If the condition \eqref{eq:np_cond} is met,
the Higgs particles suddenly acquire the following number density after the passage of the non-adiabatic region,
$|\sigma| < [m_{\text{eff},\sigma} \bar\sigma/ \tilde g]^{1/2}$:
\begin{align}
	n_H \simeq \frac{k_\ast^3}{2 \pi^3}.
\end{align}

\begin{figure}[t]
\centering
\subfigure{
\includegraphics[width=0.49\columnwidth,clip]{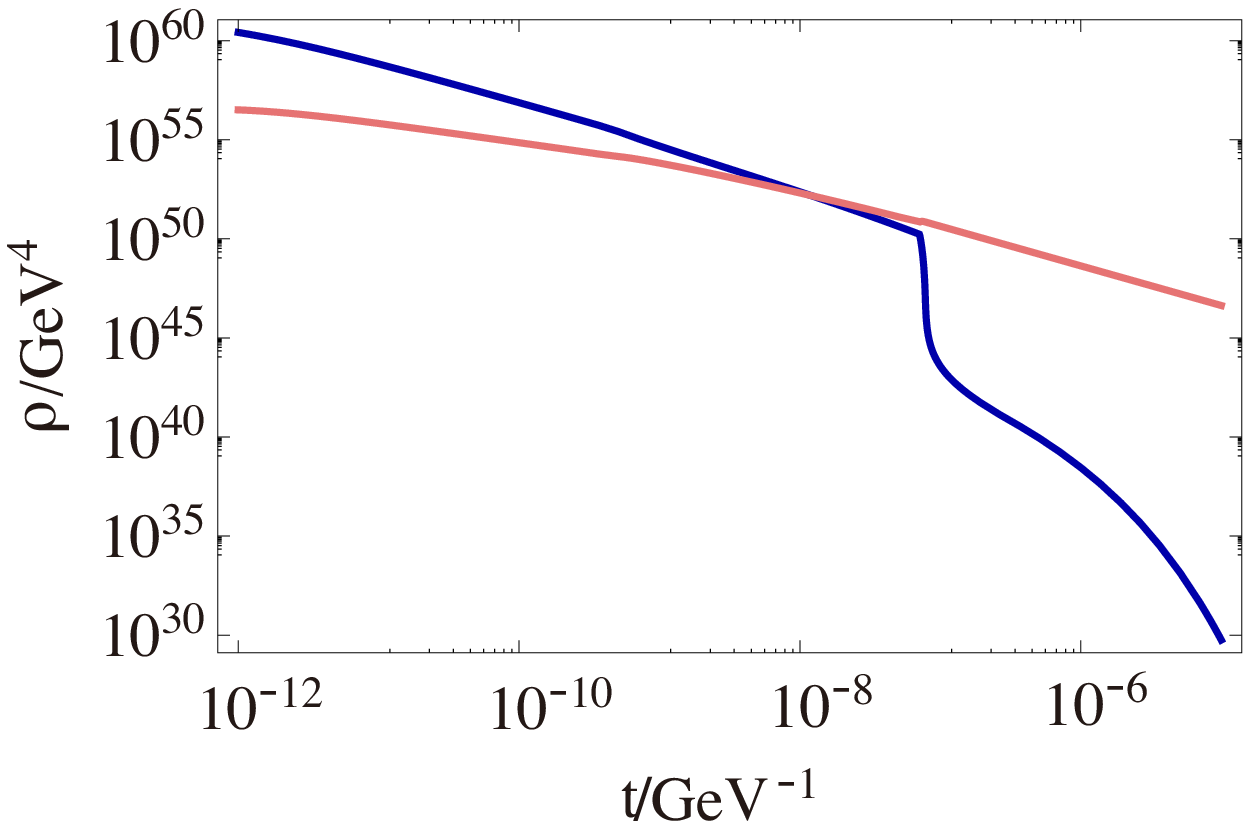}}
\subfigure{
\includegraphics[width=0.49\columnwidth,clip]{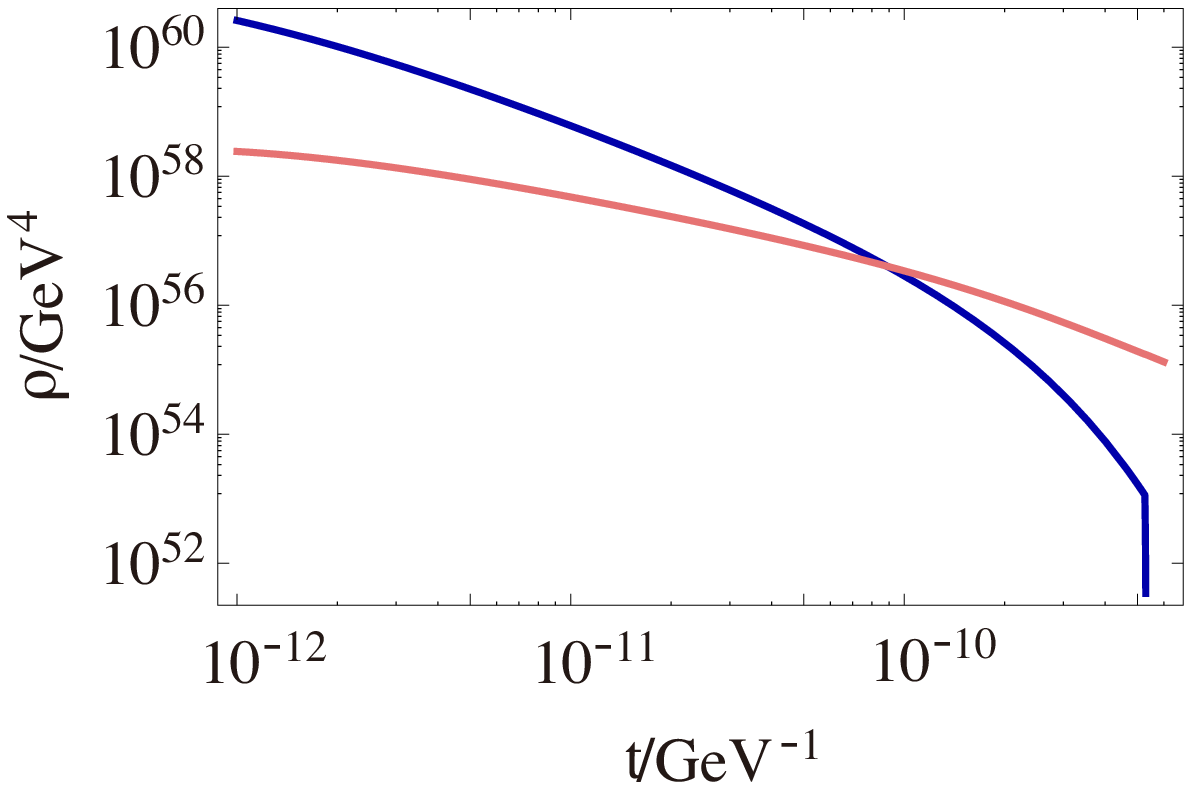}}
\subfigure{
\includegraphics[width=0.49\columnwidth,clip]{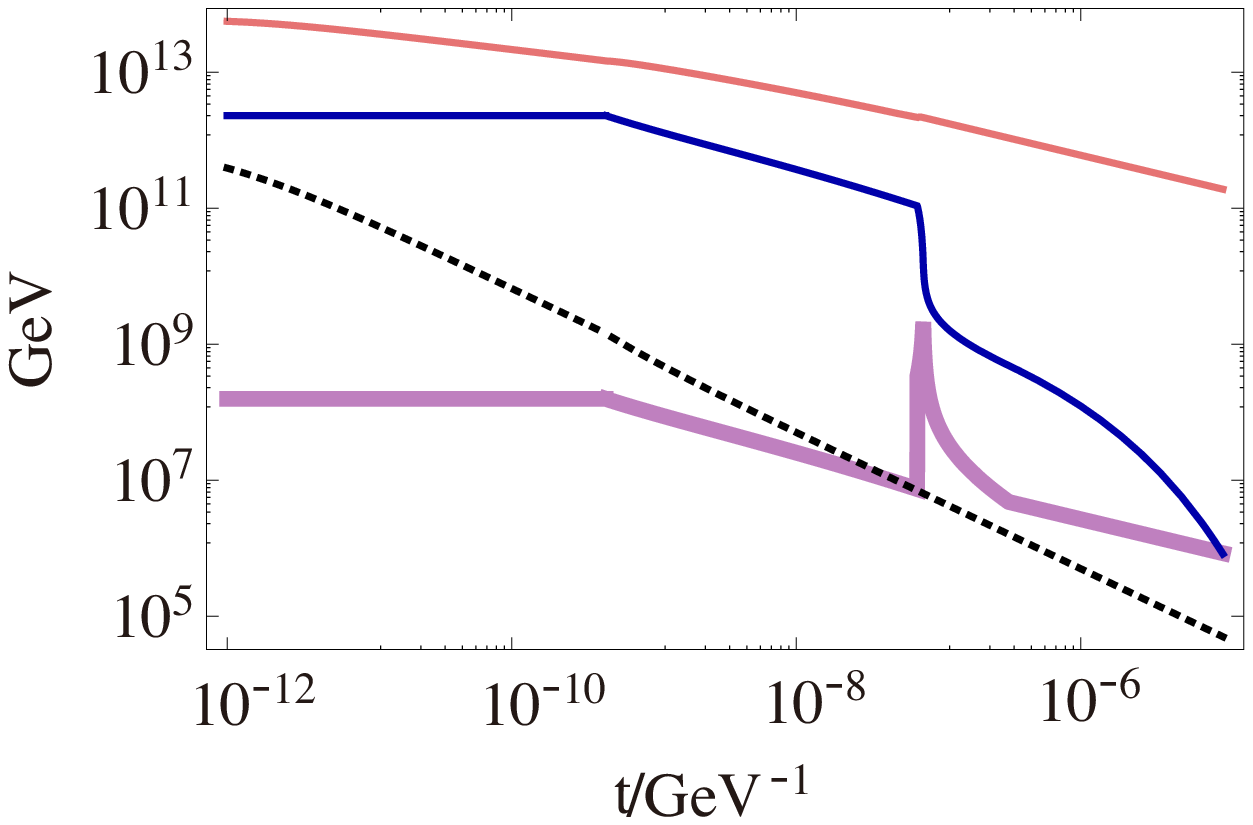}}
\subfigure{
\includegraphics[width=0.49\columnwidth,clip]{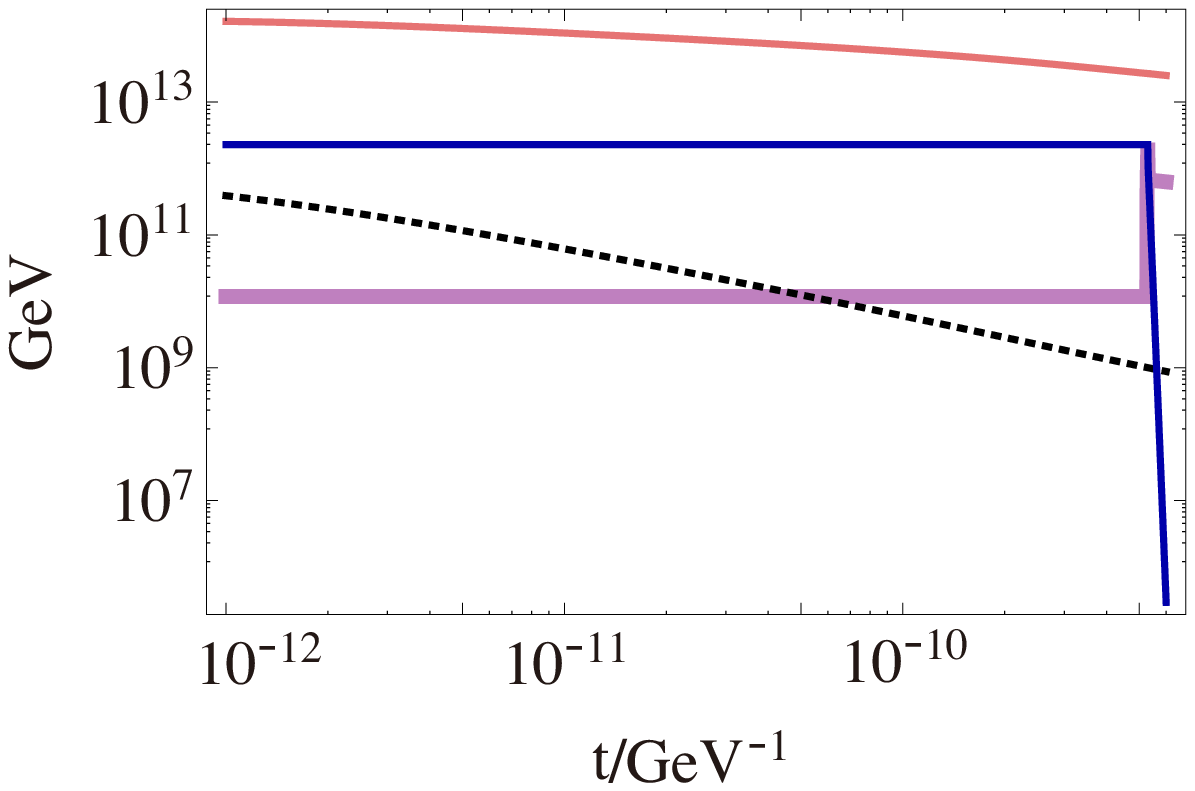}}
\caption{\small \textbf{Left Two Panels} show the typical time evolution in the light singlet case,
$(\tilde g^2, \tilde m_\sigma )= (4 \times 10^{-3},55\, \GEV)$,
\textbf{Right Two Panels} show the typical time evolution in the heavy singlet case, 
$(\tilde g^2, \tilde m_\sigma )= (0.3,1\, \TEV)$ [See Eq.~\eqref{eq:bmp}].
\textbf{Upper Two Panels} show the time evolution of the energy density for the inflaton [{\color{blue}blue}] and
for the radiation [{\color[rgb]{1.000000,0.400000,0.400000}pink}]. 
\textbf{Lower Two Panels} show the time evolution of various quantities;
the effective dissipation rate [{\color[rgb]{0.800000,0.400000,1.000000}purple}], 
the Hubble parameter [black dotted], the effective mass of the inflaton [{\color{blue}blue}]
and the temperature [{\color[rgb]{1.000000,0.400000,0.400000}pink}].}
\label{fig:ev}
\end{figure}

Since the Higgs field has a large top Yukawa coupling,
these produced Higgs particles can decay into other SM particles.
After the Higgs production at the non-adiabatic region,
the Higgs particles become heavier because the field value of the inflaton grows
towards $\bar \sigma$.
Hence, the decay rate of the Higgs into other SM particles becomes larger correspondingly.
The typical time scale can be estimate as 
$1 \sim \Gamma_H (\sigma(t_\text{dec})) t_\text{dec} \sim y_t^2 \tilde g \bar\sigma m_{\text{eff},\sigma} t_\text{dec}^2$.
If this time scale $t_\text{dec}$ is much shorter than the oscillation period of the inflaton,
$m_{\text{eff},\sigma} \ll y_t^2 \tilde g \bar\sigma \sim \tilde g \bar \sigma$,
produced Higgs particles completely decay before the inflaton moves back to its origin again.
As a result, the parametric resonance does not occur since the induced emission is absent.
In our case, this condition is almost automatically satisfied when the non-perturbative particle production
can take place,
because the top Yukawa coupling is large.
Through this process, the inflaton loses its energy with a rate~\cite{Mukaida:2012qn}:
\begin{align}
	\Gamma_\sigma^\text{NP} \simeq \frac{\tilde g^2 m_{\text{eff},\sigma}}{\pi^4 y_t}.
\end{align}
The background plasma is produced gradually by this process, and 
typically the non-perturbative particle production becomes less effective
when the screening mass of the Higgs becomes comparable to the characteristic scale,
$m_{\text{scr},H} \sim k_\ast$.
If the background plasma can be well approximated with the thermal plasma,\footnote{
	In our case, the inflaton mass
	is much smaller than the would-be temperature of the background plasma.
	Hence, we expect that this assumption is (marginally) satisfied.
	See also Fig.~\ref{fig:ev}.
}
this condition indicates $\rho_\text{rad} \sim (\tilde g^2 / y_t^4) \rho_\sigma$,
and hence the energy is still dominated by the inflaton.
And also the temperature tends to be larger than the effective mass of the inflaton
at that time, $T \gg m_{\text{eff},\sigma}$ (See also Fig.~\ref{fig:ev}).

After the end of the preheating stage,
the inflaton dominantly dissipates its energy via frequent interactions with 
abundant thermal plasma.
The thermal dissipation rates of the Z$_2$-symmetric scalar field were studied in detail in Ref.~\cite{Mukaida:2013xxa},
and so we do not repeat technical details.
Instead, let us summarize the relevant dissipation rates and explain intuitive reasoning of their behavior.
Since the oscillation period is much smaller than the temperature,
the background plasma essentially feels the inflaton as a slowly varying object.
In addition, the Higgs particles are expected to be produced from the thermal plasma
when the inflaton passes through $\tilde g |\sigma| < T$.
This is because the time interval $\delta t$ in which the inflaton passes through $\tilde g |\sigma| < T$ is
much longer than the typical production rate of the Higgs in this regime:
$y_t^2 T \delta t \sim y_t^2 T^2 / k_\ast^2 \gtrsim 1$ because $m_{\text{scr},H} \sim y_t T$.
Therefore, one can estimate the dissipation rate of the inflaton by assuming that the background plasma
remains in thermal equilibrium.
There are three typical situations where the thermal dissipation becomes relevant~\cite{Mukaida:2013xxa}.
\begin{description}
	\item[(i).~Dissipation via thermally populated Higgs particles with $m_{\text{eff},\sigma} \ll y_t^2 T$:]	
	Together with the above discussion, the inflaton can be regarded as a slowly moving object in this case.
	The inflaton loses its energy via interactions with thermally populated Higgs particles while
	it passes through $\tilde g |\sigma| < T$.
	The effective dissipation rate can be estimated as
	\begin{align}
	\label{eq:diss_slow}
		\Gamma_\sigma^\text{eff, slow} \sim c
		\begin{cases}
			\cfrac{\tilde g T^2}{y_t^2 \bar\sigma} &\text{for}~~ T /\tilde g \ll \bar \sigma , \\[15pt]
			\cfrac{\tilde g^4 \bar \sigma^2}{y_t^2T} &\text{for}~~ T < \bar \sigma \ll T/ \tilde g , \\[15pt]
			\cfrac{\tilde g^4 T}{y_t^2} &\text{for} ~~ \bar \sigma < T,
		\end{cases}
	\end{align}
	with $c$ being a factor of $\mathcal O (10^{-1})$.
	\item[(ii).~Dissipation via thermally populated Higgs particles with $\tilde g \bar\sigma \ll y_t T$:]
	In this case, one can safely assume that the Higgs particles are in the thermal plasma even if
	the inflaton oscillates fast because the amplitude $\bar \sigma$ can be neglected.
	Again, the inflaton loses its energy via interactions with thermally populated Higgs particles,
	and its rate can be estimated as
	\begin{align}
		\Gamma_\sigma^\text{eff, small} &\sim c
		\begin{cases}
			\cfrac{\tilde g^4 \bar \sigma^2}{y_t^2 T} 
			&\text{for}~~\sqrt{\frac{y_t^2 T}{ m_{\text{eff},\sigma}}} T <\bar \sigma < y_t T /\tilde g \\[15pt]
			\cfrac{\tilde g^4 T^2}{m_{\text{eff}, \sigma}}
			&\text{for} ~~ \bar \sigma < \sqrt{\frac{y_t^2 T}{ m_{\text{eff},\sigma}}} T
		\end{cases}
		&\text{with}~~ y_t^2 T \ll m_{\text{eff}, \sigma} \lesssim y_t T; \\[20pt]
		\Gamma_\sigma^\text{eff, small}
		&\sim c
		\begin{cases}
			\cfrac{\tilde g^4 \bar\sigma^2}{m_{\text{eff}, \sigma}}
			&\text{for}~~ T < \bar\sigma < y_t T / \tilde g \\[15pt]
			\cfrac{\tilde g^4 T^2}{m_{\text{eff},\sigma}}
			&\text{for}~~ \bar \sigma < T
		\end{cases}
		&\text{with}~~ y_t T \lesssim m_{\text{eff}, \sigma} < T; \\[20pt]
		\Gamma_\sigma^\text{eff, small}
		&\sim c \frac{\tilde g^4 \bar \sigma^2}{m_{\text{eff}, \sigma}}
		&\text{with}~~  T < m_{\text{eff},\sigma},
	\end{align}
	with $c$ being a factor of $\mathcal O (10^{-1})$.
	\item[(iii).~Dissipation via a higher dimensional operator with $m_{\text{eff},\sigma} \ll y_t^2 T$:]
	In the regime where the field value of the inflaton is larger than $\tilde g |\sigma| > T$,
	there are no Higgs particles due to the Boltzmann suppression.
	Even in this regime,
	integrating out these heavy Higgs field, one finds that the inflaton can interact with
	the thermal plasma via a higher dimensional operator:
	$(\delta \sigma / \bar \sigma) FF$,
	which leads to the following effective dissipation rate~\cite{Bodeker:2006ij,Laine:2010cq}:
	\begin{align}
		\Gamma_\sigma^\text{eff, large} \sim b \alpha^2 \tilde g  \frac{T^2}{\bar \sigma},
	\end{align}
	with $b$ being a factor of ${\cal O} (10^{-3})$ and $\alpha$ being the fine structure constant
	of SU(2)$_\text{W}$. 
\end{description}
Note that all the above dissipation rates contain uncertainties $c$
that come from approximations used in the course of calculation~\cite{Mukaida:2012qn}.
In the following, we do not seriously care about factors but concentrate on the order of magnitude estimation.

In order for the inflaton condensation to lose its energy completely,
its dissipation rate should exceed the Hubble parameter before $\tilde m_\sigma > y_t^2 T$.
Otherwise the dissipation rate cannot catch up the Hubble expansion.
This puts a lower bound on the coupling $\tilde g \gtrsim \tilde g_c \equiv [\tilde m_\sigma / \mpl]^{1/4}$,
which is satisfied in the relevant parameter spaces.
After the inflaton coherent oscillation loses its energy completely,
then the inhomogeneous mode of the inflaton is expected to cascade towards the ultra-violet regime, $T$, immediately against the cosmic expansion 
and eventually participates in the thermal bath~\cite{Mukaida:2013xxa}.\footnote{
	Also one can show that the preheating due to the four point self-interaction of the inflaton
	does not change our results quantitatively because $\tilde \lambda\sim \tilde g^4 \ll \tilde g^2$.
	See Ref.~\cite{Mukaida:2014yia} for more detailed discussion.
}

Now we are in a position to estimate the reheating temperature.
To be concrete, we perform numerical calculation of Eqs.~\eqref{eq:eom1}--\eqref{eq:eom3}
with two benchmark points: the light/heavy singlet case~\eqref{eq:bmp}.
Fig.~\ref{fig:ev} shows the time evolution of various quantities in the two cases:
Left/Right two panels for the Light/Heavy cases.
The upper two panels show the evolution of the energy density of inflaton and radiation with
blue and pink solid lines respectively.
And the lower two panels show the evolution of the effective dissipation rate [purple solid],
the Hubble parameter [black dotted], the effective mass of the inflaton [blue solid] and
the temperature [pink solid].
In both cases, at first, the radiation is produced via the instant preheating and it behaves as
$\rho_\text{rad} \sim \Gamma^\text{NP}_\sigma \rho_\sigma /H \propto m_{\text{eff}, \sigma} \rho_\sigma^{1/2}$
until $k_\ast \sim y_t T$ where the condition \eqref{eq:np_cond} is broken down.\footnote{
	In the light singlet case,
	at the very time when the non-perturbative particle production shuts off,
	the energy densities of the inflaton and radiation become comparable,
	since the condition $y_t T \simeq k_\ast$ [See Eq.~\eqref{eq:np_cond}] 
	implies $\rho_\text{rad}/\rho_\sigma \simeq \pi^2 g_\ast \tilde g^2/(30 y_t^4 ) \sim 1$
	accidentally.
}
In the light singlet case (left panels), the energy density of radiation scales as $\rho_\text{rad} \propto a^{-3/2}$
when the inflaton oscillates with the quadratic term,
and then the scaling changes as $\rho_\text{rad} \propto a^{-3}$ since the inflaton potential 
becomes dominated by the quartic term.
When the condition \eqref{eq:np_cond} is saturated,\
the thermal dissipation comes in and the amplitude of the inflaton immediately decreases
$\tilde g \bar\sigma < T$.
Owing to the thermally populated Higgs particles, the dissipation rate suddenly increases
and eventually the inflaton condensation is completely broken into particles by the dissipation
$\Gamma^\text{eff, slow}_\sigma \propto T$ in the last line of Eq.~\eqref{eq:diss_slow}.\footnote{
	The evolution of the Universe after the reheating $t \gg 10^{-7} \, \GEV^{-1}$ strongly depends
	on the uncertainty $c$,  but its behavior is qualitatively correct.
}
In the heavy singlet case (right panels), the Universe is soon dominated by the radiation via
the instant preheating before the potential is dominated by the quartic term.
After the condition \eqref{eq:np_cond} is saturated, 
the inflaton is soon dissipated away due to the thermally populated Higgs particles.

In both cases, the radiation starts to dominate the Universe at $H \sim \Gamma^\text{NP}_\sigma$,
which indicates the reheating temperature: 
$T_\text{R} \sim [90/(\pi^2 g_\ast)]^{1/4} \sqrt{\Gamma^\text{NP}_\sigma \mpl}$
with $g_\ast$ being the relativistic degree of freedom at $T_\text{R}$.
It can be expressed as 
\begin{align}
	T_\text{R} \sim \left( \frac{90}{\pi^2 g_\ast} \right)^{1/4}
	\begin{cases}
	 	\left( \cfrac{3}{2 \pi^{18}} \right)^{1/4} \left( \cfrac{\tilde g^3}{y_t} \right) \mpl &\text{for light singlet case}, \\[20pt]
		\left(\cfrac{\tilde g}{\pi^2 \sqrt{y_t}} \right) \sqrt{M \mpl} &\text{for heavy singlet case}.
	 \end{cases}
\end{align}
As one can see, the radiation dominant era starts much before $T \sim 10^{12}\, \GEV$,
even in the light singlet case where the interaction $\tilde g$ is relatively small.

\section{Discussion and Conclusions}
\label{sec:}

Motivated by the recent observation of the B-mode polarization by the BICEP2 experiment,
we have considered a scenario that the chaotic inflation is induced by a singlet scalar field with the running kinetic term,
and it simultaneously becomes DM in the present Universe.
The key point is that the inflaton is heavy at the large field value where inflation happens,
while it can be so light around the potential minimum that its thermal relic abundance can match with observed DM abundance.

The process of reheating might be non-trivial since the inflaton has a Z$_2$-symmetry and it is perturbatively stable at its vacuum.
However, the combined effects of particle production and the scattering with particles in thermal bath
cause efficient dissipation on the inflaton coherent oscillation and actually the inflaton is soon thermalized after inflation ends.
Once the inflaton participates in the thermal plasma, the following thermal history is described by the standard radiation dominated Universe.
Since the reheating temperature is so high, $T_\text{R} \gtrsim 10^{12}\, \GEV$, the thermal leptogenesis successfully works~\cite{Fukugita:1986hr}.
Interestingly, the four point interaction with the SM Higgs that determines the DM relic density
also induces the cosmic reheating and determines the reheating temperature.
Thus we think that this is a kind of minimal scenario that explains the primordial inflation, reheating and present DM abundance.

The singlet scalar DM scenario may be probed by future direct DM detection experiments~\cite{Aprile:2012zx}
and also by the collider searches through the Higgs invisible decay~\cite{Peskin:2012we}.
Because of the high inflation scale and high reheating temperature, direct detection of inflationary gravitational waves
with future space laser interferometers is also plausible~\cite{Turner:1990rc,Seto:2003kc,Smith:2005mm,Nakayama:2008wy}.

\section*{Acknowledgment}
This work is supported by Grant-in-Aid for Scientific
research from the Ministry of Education, Science, Sports, and Culture
(MEXT), Japan.
The work of K.M. is supported in part by JSPS Research Fellowships
for Young Scientists.
This work was supported by the Grant-in-Aid for Scientific Research on
Innovative Areas (No. 21111006  [KN]), Scientific Research (A) (No. 22244030 [KN]).




\end{document}